\documentclass[12pt, appendixfloats, numberedappendix]{emulateapj}
\usepackage{amsmath}
\usepackage{color}
\usepackage[normalem]{ulem}

\def\eg{e.g., }

\newcommand{\mpc}{{\rm\,Mpc}}
\newcommand{\kpc}{{\rm\,kpc}}
\newcommand{\pc}{{\rm\,pc}}

\newcommand{\beq}{\begin{equation}}
\newcommand{\eeq}{\end{equation}}

\def\nii{[\ion{N}{2}]}

\def\oiii{[\ion{O}{3}]}

\def\ha{H$\alpha$}

\newcommand{\kms}{\ensuremath{{\rm km\,s}^{-1}}}

\newcommand{\MSun}{\ensuremath{\rm M_\odot}}

\newcommand{\ud}{\ensuremath{{\rm d}}}

\usepackage{natbib}
\usepackage{rotating}
\begin{document}

\shorttitle{Leaky-box Model for the Local Relation}
\shortauthors{Zhu et al.}
\title{A Local Leaky-box Model for the Local Stellar Surface Density - Gas Surface Density - Gas Phase Metallicity Relation}

\author{
Guangtun~Ben~Zhu\altaffilmark{1,2}, Jorge~K.~Barrera-Ballesteros\altaffilmark{1},
Timothy~M.~Heckman\altaffilmark{1}, Nadia~L.~Zakamska\altaffilmark{1,3}, \\
Sebastian~F.~S\'{a}nchez\altaffilmark{4},
Renbin~Yan\altaffilmark{5}, 
Jonathan~Brinkmann\altaffilmark{6}
}
\altaffiltext{1}{Department of Physics \& Astronomy, Johns Hopkins University, 3400 N. Charles Street, Baltimore, MD 21218, guangtun.ben.zhu@gmail.com}
\altaffiltext{2}{Hubble Fellow}
\altaffiltext{3}{Deborah Lunder and Alan Ezekowitz Founders' Circle Member, Institute for Advanced Study, Einstein Dr., Princeton, NJ 08540, USA}
\altaffiltext{4}{Instituto de Astronom\'{i}a, Universidad Nacional Aut\'{o}noma de M\'{e}xico, A.P. 70-264, 04510 M\'{e}xico, D.F., M\'{e}xico}
\altaffiltext{5}{Department of Physics and Astronomy, University of Kentucky, 505 Rose St., Lexington, KY 40506-0057}
\altaffiltext{6}{Apache Point Observatory, P.O. Box 59, Sunspot, NM 88349}

\begin{abstract}
We revisit the relation between the stellar surface density, the gas surface density, and 
the gas-phase metallicity of typical disk galaxies in the local Universe with the SDSS-IV/MaNGA survey, 
using the star formation rate surface density as an indicator for the gas surface density.
We show that these three local parameters form a tight relationship, confirming previous works 
(e.g., by the PINGS and CALIFA surveys), but with a larger sample.
We present a new local leaky-box model, assuming star formation history and chemical evolution is localized except for outflowing materials. 
We derive closed-form solutions for the evolution of stellar surface density, gas surface density and gas-phase metallicity, 
and show that these parameters form a tight relation independent of initial gas density and time.
We show that, with canonical values of model parameters, this predicted relation match the observed one well.
In addition, we briefly describe a pathway to improving the current semi-analytic models of galaxy formation by incorporating the local leaky-box model
in the cosmological context, which can potentially explain simultaneously multiple properties of Milky Way-type disk galaxies, 
such as the size growth and the global stellar mass-gas metallicity relation.


\end{abstract}


\section{Introduction}\label{intro}



Over the past few decades, a standard cosmological model of structure formation emerged 
in a series of major observational and theoretical advances \citep[\eg][]{white78a}. 
However, most of these studies have largely focused on the global properties of galaxies
\citep[\eg][]{kauffmann93a, springel05a, somerville15a}.

Recent integral-field-unit (IFU) spectroscopic surveys from the ground \citep[\eg][]{bacon01a, rosales10a, sanchez12a},
high-spatial resolution deep imaging surveys with the Hubble Space Telescope \citep[\eg][]{scoville07a, koekemoer11a}, 
and high-resolution hydrodynamical simulations \citep[\eg][]{vogelsberger14a, hopkins14a} have shifted the focus of 
the investigations of galaxy formation to small-scale astrophysics and to the relationships between local and global properties of galaxies.
In particular, the MaNGA survey \citep[][]{bundy15a} in SDSS-IV \citep[][]{blanton17a} is
obtaining IFU spectroscopy for about $10,000$ nearby galaxies and will provide the largest sample of galaxies with kpc-scale 
resolved optical spectroscopy, enabling systematic investigations of local properties and also their correlations with global parameters.
In this paper, using the MaNGA data obtained in the first two years, we investigate the relation 
between the stellar surface density ($\Sigma_*$), gas surface density ($\Sigma_{\rm gas}$), and gas-phase metallicity
($Z$) in typical disk galaxies,
using the star formation rate (SFR) surface density ($\Sigma_{\rm SFR}$) as a proxy for $\Sigma_{\rm gas}$. 
In particular, we show that a simple leaky-box model can explain well the observed relation between these parameters and propose
a new way of thinking about disk galaxy formation.

The rest of the paper is organized as follows. In Section~\ref{sec:data} and \ref{sec:relation}, we describe the data we use and the observed relation. 
We present the local leaky-box model in Section~\ref{sec:leakymodel}. In Section~\ref{sec:insideout}, we outline a global semi-analytic model for disk galaxy formation. 
We summarize our results in Section~\ref{sec:summary}.
When necessary, we assume the $\Lambda$CDM cosmogony, with $\Omega_\Lambda=0.7$, $\Omega_{\mathrm m}=0.3$, and ${\mathrm H}_0=70\,\kms\,\mpc^{-1}$. 

\begin{figure*}
\epsscale{0.55}
\plotone{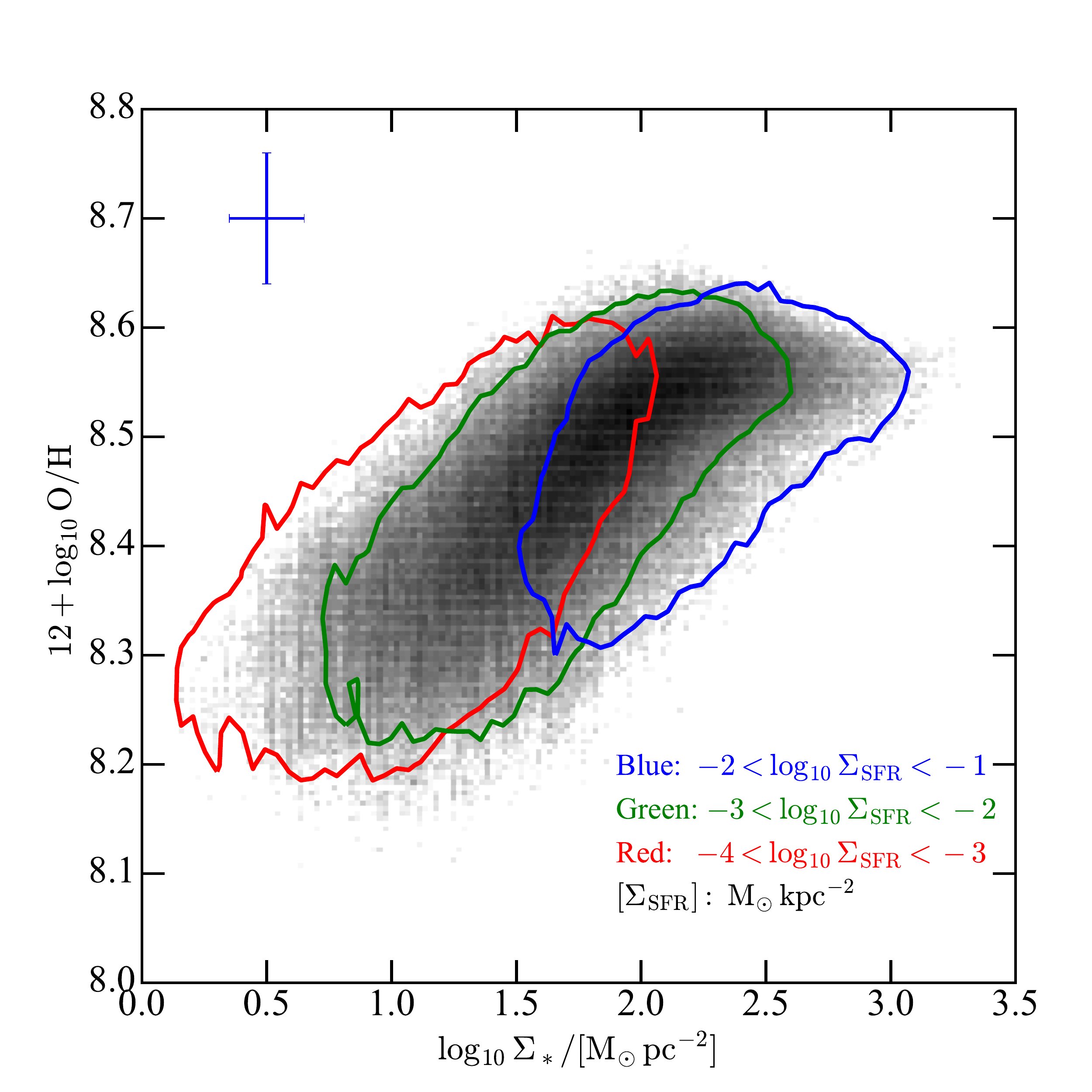}
\epsscale{0.55}
\plotone{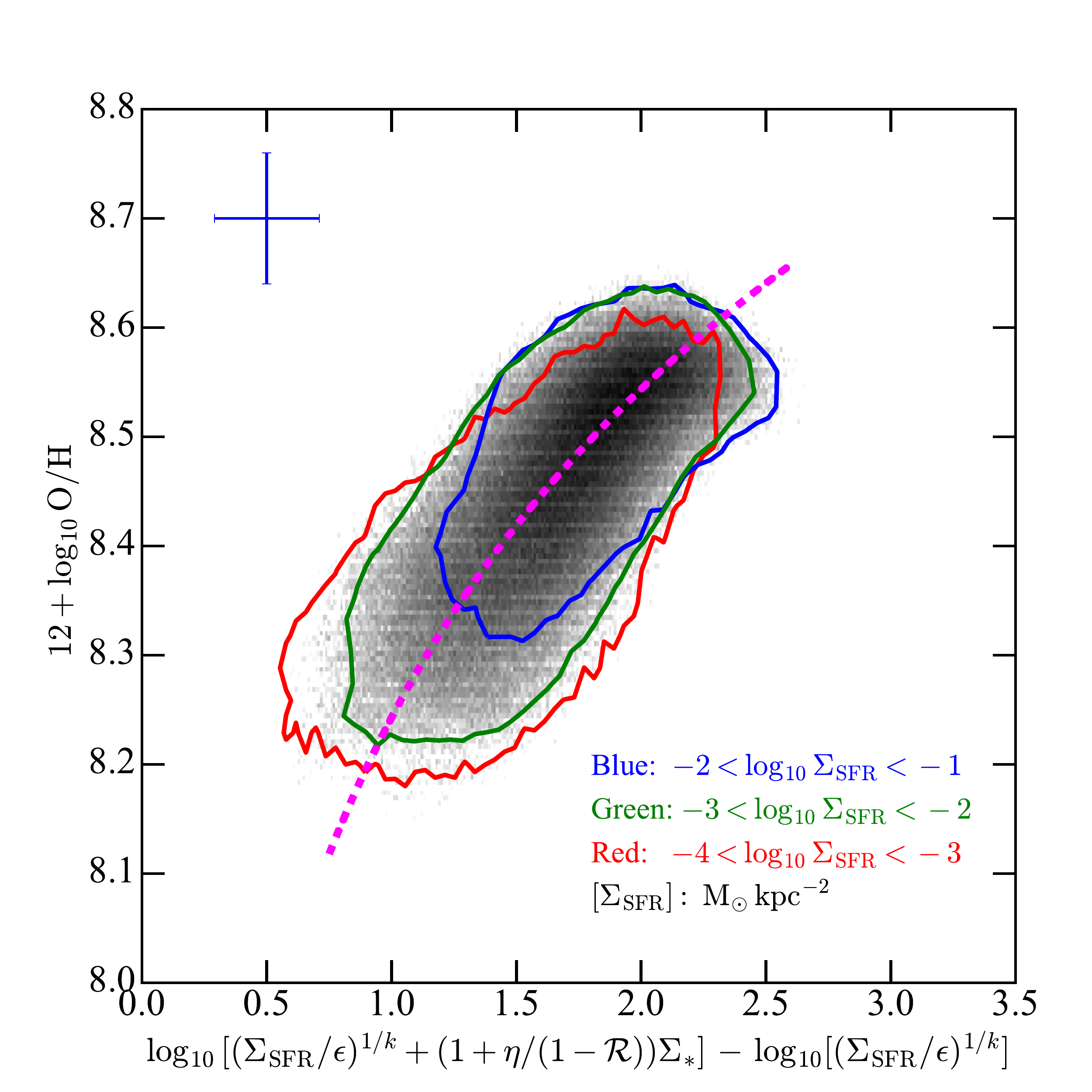}
\caption{\textit{Left}: The observed $\Sigma_{*}-Z$ relation of star-forming regions in typical disk galaxies. 
The contours enclose $90\%$ of the subsamples with highest (blue), intermediate (green) and lowest (red) SFR surface density. 
\textit{Right}: The observed $\Sigma_{*}-\Sigma_{\rm SFR}-Z$ relation (gray scale, Eq.~\ref{eq:metalgas}),
assuming $\mathcal{R}=0.3$, $\epsilon=0.0004$, $k=2.2$, and $\eta=1$. The dashed line shows the relation with the 
best-fit yield $y=0.003$. The errorbars show the typical measurement uncertainties,
$0.06\,$dex for metallicity and $0.15\,$dex for stellar and SFR surface density.
}
\vspace{0.2cm}
\label{fig:relation}
\end{figure*}

\section{Data}\label{sec:data}
The SDSS-IV/MaNGA IFU survey uses the BOSS spectrographs \citep[]{smee13a} 
on the 2.5-m SDSS telescope \citep[][]{gunn06a} at the Apache Point Observatory.
Detailed description of the MaNGA surveys are available in \citet[][overview]{bundy15a}, \citet[][instrumentation]{drory15a},
\citet[][observation, data reduction]{law15a, law16a}, and \citet[][calibration, survey design]{yan16a, yan16b}.
We use the fourth internal data release of the MaNGA survey (MPL-4), which includes $1390$ galaxies observed as of June 2015. 

For our purposes, we are interested in typical disk galaxies and we select our sample and use the same data as we did
in \citet{barrera16a}. We select $653$ disk galaxies spanning stellar masses between $10^{8.5}\,\MSun$ and
$10^{11}\,\MSun$. The data cubes include about $507,000$ star-forming spaxels with spatial resolution ranging 
from $\sim1.5\,\kpc$ to $\sim2.5\,\kpc$.
For the parameter measurements, we use the estimates from the PIPE3D pipeline \citep{sanchez16a}.
PIPE3D estimated the stellar mass at a given spaxel by fitting the underlying stellar continuum with spectral templates taken 
from MIUSCAT SSP library \citep[][]{vazdekis12a}, assuming a \citet{salpeter55a} IMF. 
The pipeline also took into account of dust attention \citep[][]{calzetti01a}.
We estimated SFR using the dust attenuation-corrected flux of \ha. 
We have also corrected the surface densities for the inclination effect \citep[see][]{barrera16a}.
For gas-phase metallicity, we use the O3N2 indicator based on the \oiii$\,\lambda5008$ and \nii$\,\lambda6584$ ratio \citep[\eg][]{marino13a}.
For more details regarding the data and the survey, we refer the reader to references above.

\section{The local $\Sigma_{*}-\Sigma_{\rm SFR}-Z$ relation}\label{sec:relation}

Early works \citep[\eg][]{edmunds84a, vilacostas92a} have already suggested that there exists a relationship between the local stellar surface density
and the gas-phase metallicity. More recently, the PINGS and CALIFA surveys have presented conclusive evidence for such a relationship \citep{rosales12a, sanchez13a}.
In \citet{barrera16a}, we presented further evidence with the MaNGA survey. 
\citet{rosales12a} and \citet{sanchez13a} further showed that, including the local SFR surface density indicates that the three parameters together form a tight relationship.
Our objective is to revisit this relation with a larger sample and then devise a local chemical evolution model for its interpretation.



In the left panel of Figure~\ref{fig:relation}, we show the $\Sigma_{*}-Z$ relation (the same as in Figure~2 of \citealt{barrera16a}). 
In addition, we divide the star-forming regions into three subsamples with the highest, intermediate, and lowest SFR surface density 
and show their distributions in blue, green, and red contours, respectively. 
We find that these three parameters, $\Sigma_{*}$, $\Sigma_{\rm SFR}$ and $Z$,  form a tight correlation with each other.
We therefore confirm the findings by \citet{rosales12a} with the PINGS survey \citep[][]{rosales10a}, 
who used luminosity surface density as a proxy for stellar surface density and \ha\ equivalent width for specific SFR, 
and also the recent results with the derived physical parameters from the larger CALIFA survey \citep[][]{sanchez13a}.


The gas-phase metallicity is the ratio of the amount of heavy elements (in our case, oxygen) to the total amount of gas in the galaxy, 
i.e., $Z=\Sigma_{\rm metal}/\Sigma_{\rm gas}$. 
Both metals and stars are integrated products of the star-formation history, while the SFR is closely correlated to 
the amount of gas available, through the Kennicutt-Schmidt (K-S) law \citep{schmidt59a, kennicutt98a}. 
The relations between the three parameters must therefore be closely related to the local star-formation history. 
In the next section, we present a leaky-box model of the local star-formation history and chemical evolution
and show that it can naturally explain our observation.


\begin{figure*}
\epsscale{0.55}
\plotone{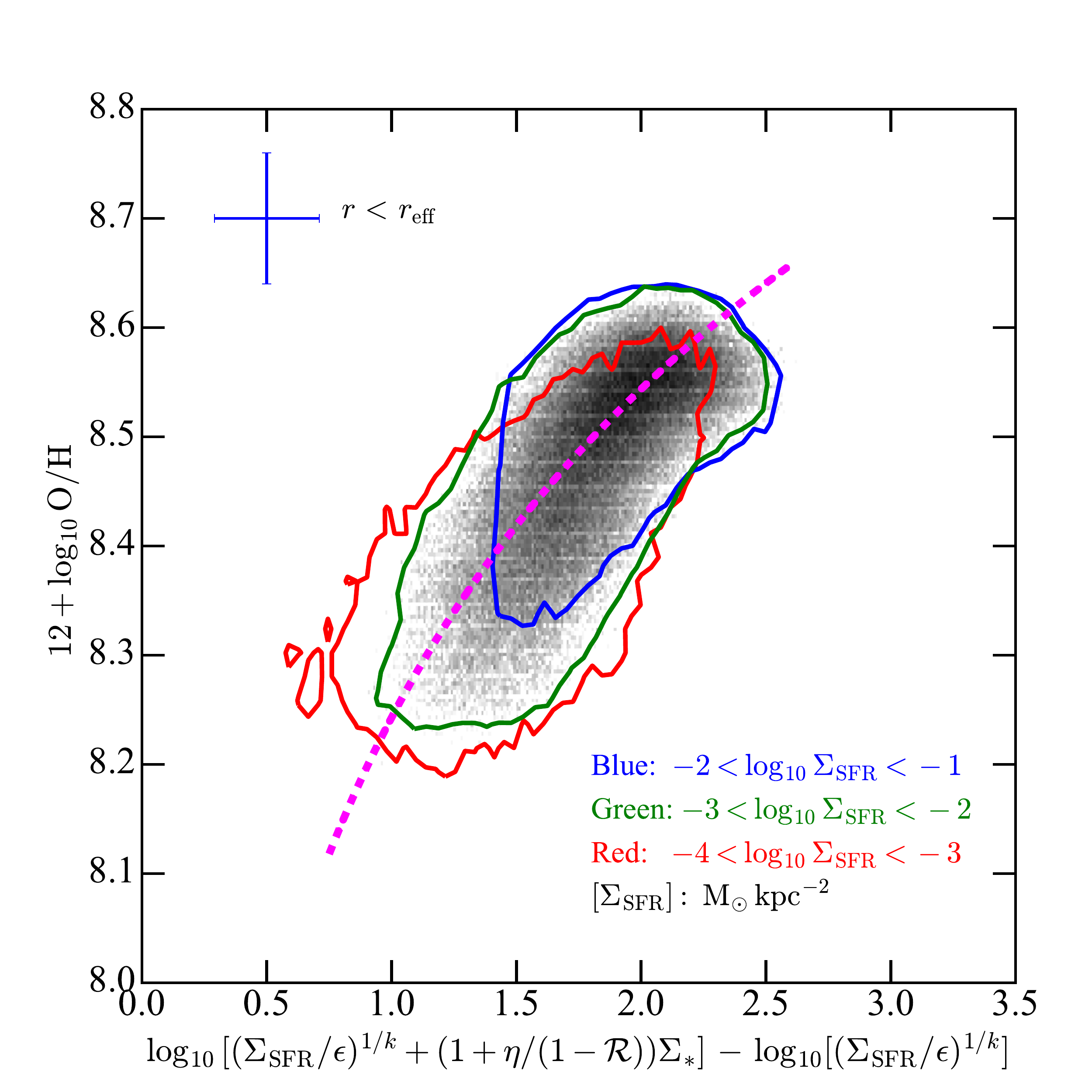}
\epsscale{0.55}
\plotone{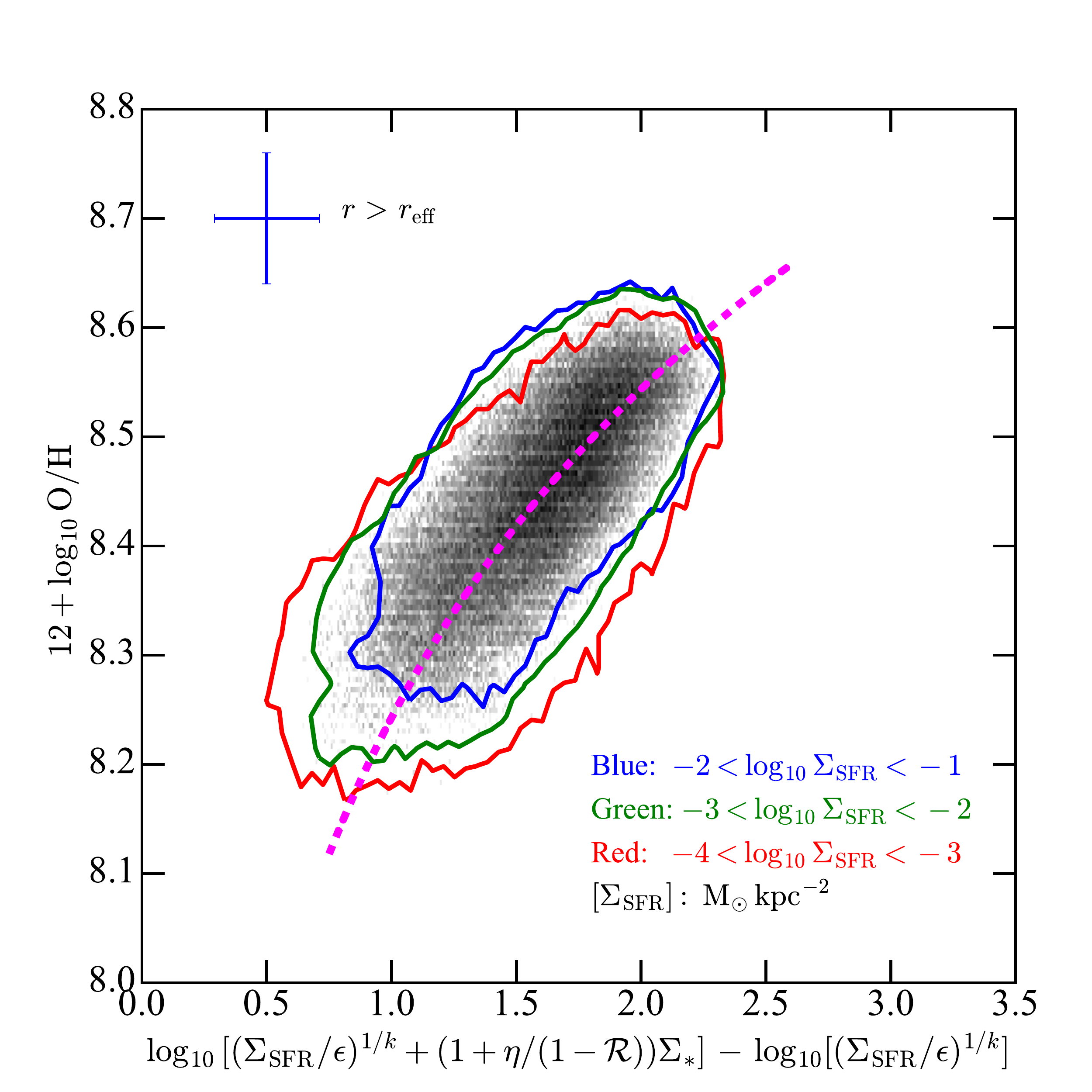}
\caption{Radial dependence of the local $\Sigma_{*}-\Sigma_{\rm SFR}-Z$ relation. \textit{Left}: Regions within $r_{\rm eff}$. \textit{Right}: Regions outside $r_{\rm eff}$. The dashed lines are the same as in Fig.~\ref{fig:relation}.
}
\vspace{0.2cm}
\label{fig:radialdependence}
\end{figure*}

\begin{figure*}
\epsscale{0.55}
\plotone{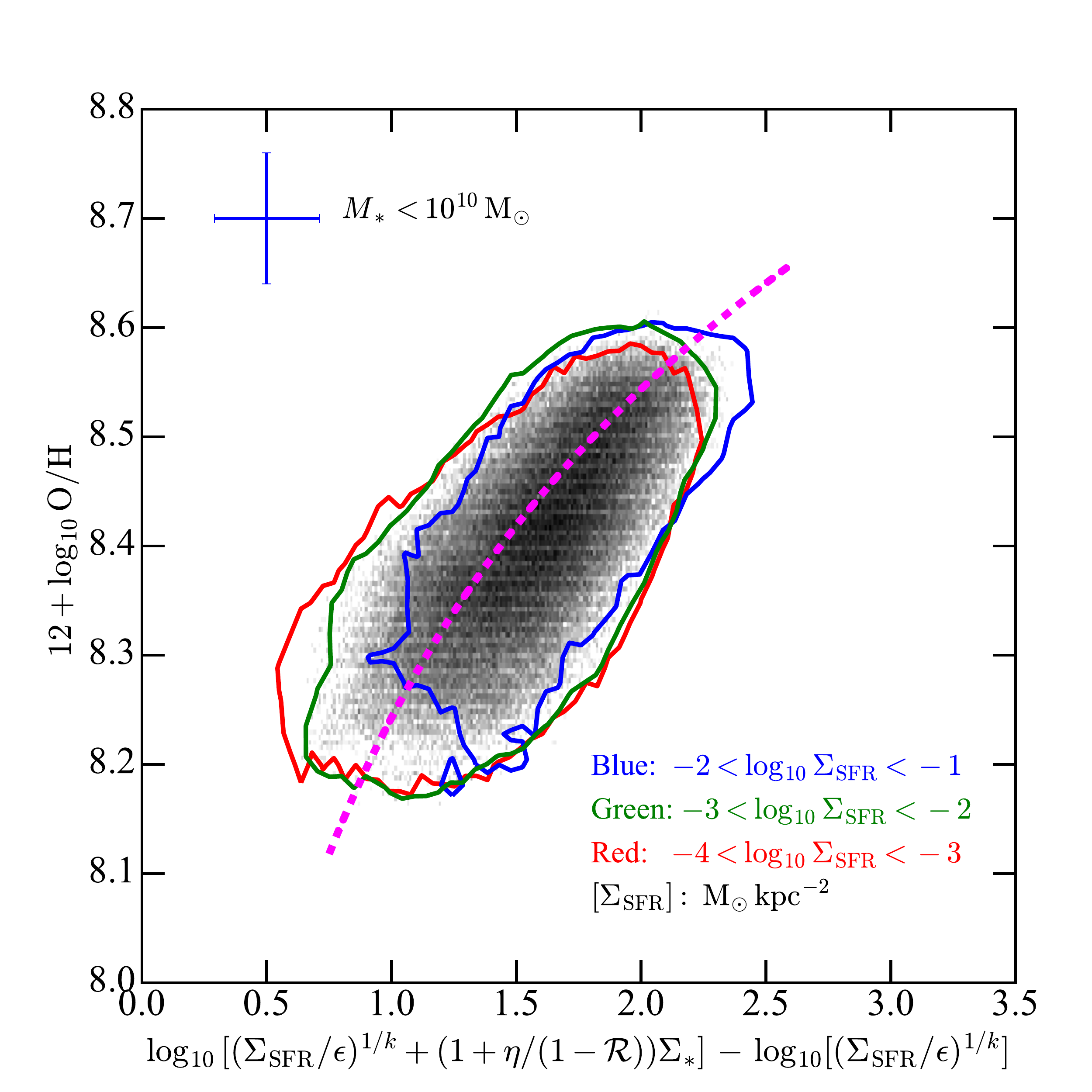}
\epsscale{0.55}
\plotone{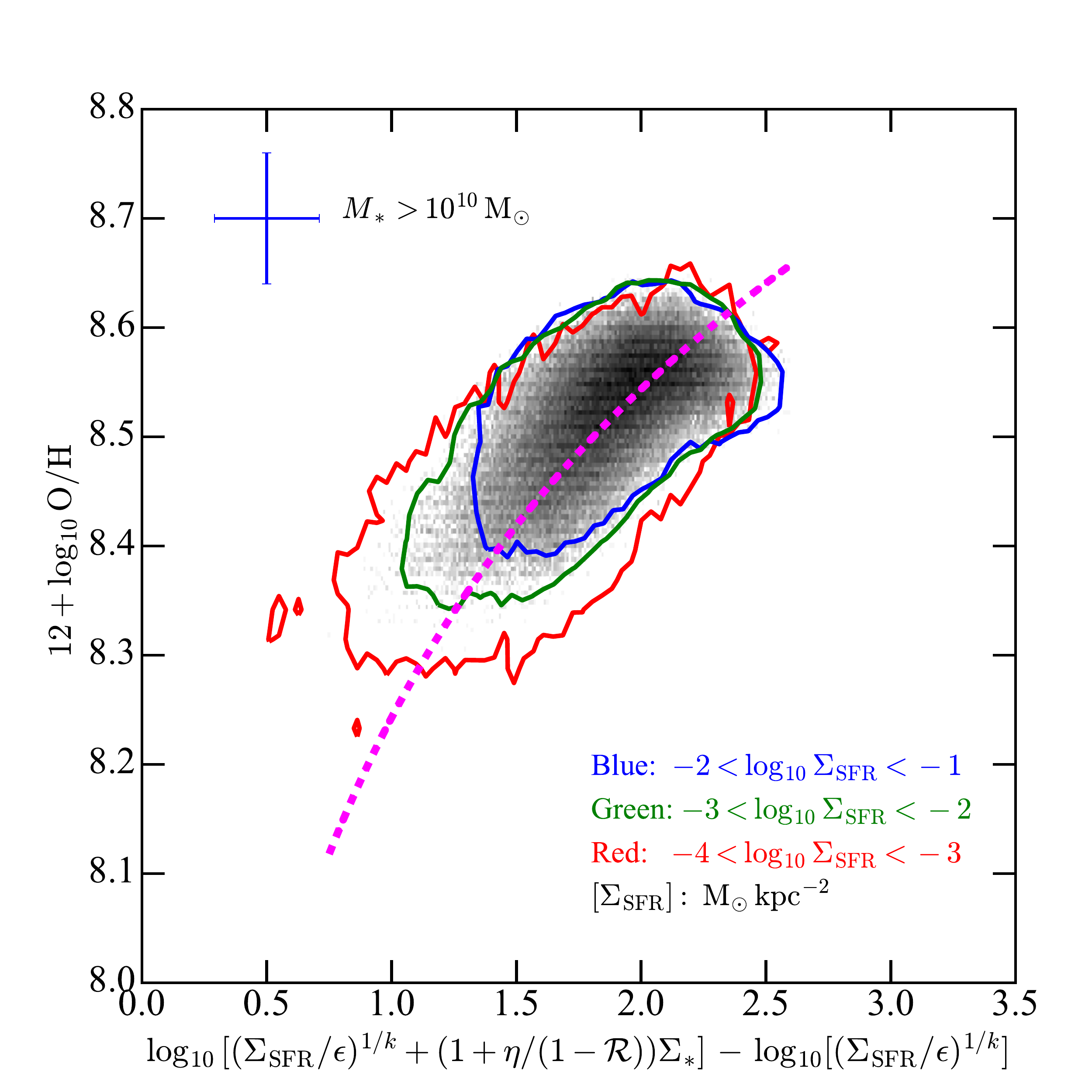}
\caption{Mass dependence of the local $\Sigma_{*}-\Sigma_{\rm SFR}-Z$ relation. \textit{Left}: Regions in host galaxies with $M_*<10^{10}\,$\MSun. \textit{Right}: Regions in host galaxies with $M_*>10^{10}\,$\MSun. The dashed lines are the same as in Fig.~\ref{fig:relation}.
}
\vspace{0.2cm}
\label{fig:massdependence}
\end{figure*}

\section{The local leaky-box model}\label{sec:leakymodel}

We assume a disk galaxy grows inside out \citep[\eg][among others]{larson76a, matteucci89a, governato07a, pilkington12a, gibson13a},
and gas falls in onto the outskirts, collapses and triggers star formation.\footnote{We note if we start with a disk of gas right from the beginning, our analysis still applies.}
In this scenario all processes -- star formation and metal production -- are localized within the same region except for the outflowing gas.
These assumptions enable us to construct a model of the localized star formation history
and chemical evolution, which we describe in detail below.

If gas is accreted onto the galaxy with initial gas surface density $\Sigma_0 \equiv \Sigma_{\rm gas}(t_0)$ at accretion time $t_0$, 
we can define a total surface density:
\begin{eqnarray}
\Sigma_{\rm tot}(t) & = & \Sigma_*(t) + \Sigma_{\rm gas}(t) + \Sigma_{\rm out}(t) \nonumber \\
          & = & \Sigma_{\rm tot} (t_0) \nonumber \\
          & = & \Sigma_0\,\mathrm{,}
\label{eq:sigma0}
\end{eqnarray}
\noindent where $\Sigma_{\rm gas}(t)$ and $\Sigma_*(t)$ are the surface densities of gas and long-lived stars at a given time $t$, respectively.
For convenience we have defined $\Sigma_{\rm out} (t)$ to represent the would-be density of the expelled gas 
should it stay within the same area, even though it can be anywhere in the circum-/inter-galactic media. 
If there is no outflow (i.e., $\Sigma_{\rm out} = 0$), we have a closed-box model. 
There has been ample evidence showing that star-forming galaxies exhibit 
ubiquitous outflows \citep[\eg][among others]{lynds63a, bland88a, heckman90a, shapley03a, rupke05a, martin09a, weiner09a, rubin14a, zhu15a}.
Outflows also help explain the large amount of metals found outside galaxies in the circum-/inter-galactic 
media \citep[\eg][among others]{bergeron86a, steidel10a, tumlinson11a, stocke13a, borthakur13a, werk14a, bordoloi14a, zhu14a}. We here therefore assume a leaky-box model.

Another assumption of our model is that the expelled gas does not fall back onto the galaxy.
Theoretical studies have suggested at least a fraction of the expelled gas would be 
reaccreted \citep[\eg][]{oppenheimer10a, bower12a, marasco12a, brook12a, henriques13a, christensen16a}.
If some of the expelled gas falls right back onto the same region, its effect is equivalent to a smaller outflow rate and our model still applies.
If some of the expelled gas gets mixed with gas outside and falls back in onto the outskirts, the formalism applies as well since the recycled gas does not invalidate the locality.
If a significant fraction of the expelled gas is spread out and falls back over the whole galaxy \citep[e.g., as in the galaxy fountain model,][]{marasco12a}, it may have a non-negligible effect 
on the chemical evolution. This last scenario is more complicated than our simple model can yet address and we leave it for future work.

With the assumptions above, the total surface density defined above stays constant over the cosmic 
time ($=\Sigma_0$). This synthetic density, $\Sigma_{\rm tot}(t)$, includes the outflowing gas, 
while the total density within the disk would only include the gas and stars in the disk ($\Sigma_*(t) + \Sigma_{\rm gas}(t)$). 
The constancy of this density and the direct connection between the amount of outflowing gas and the instantaneous SFR
make it possible to derive a closed-form solution of the full chemical evolution history, as described below.


\begin{figure}
\epsscale{1.15}
\plotone{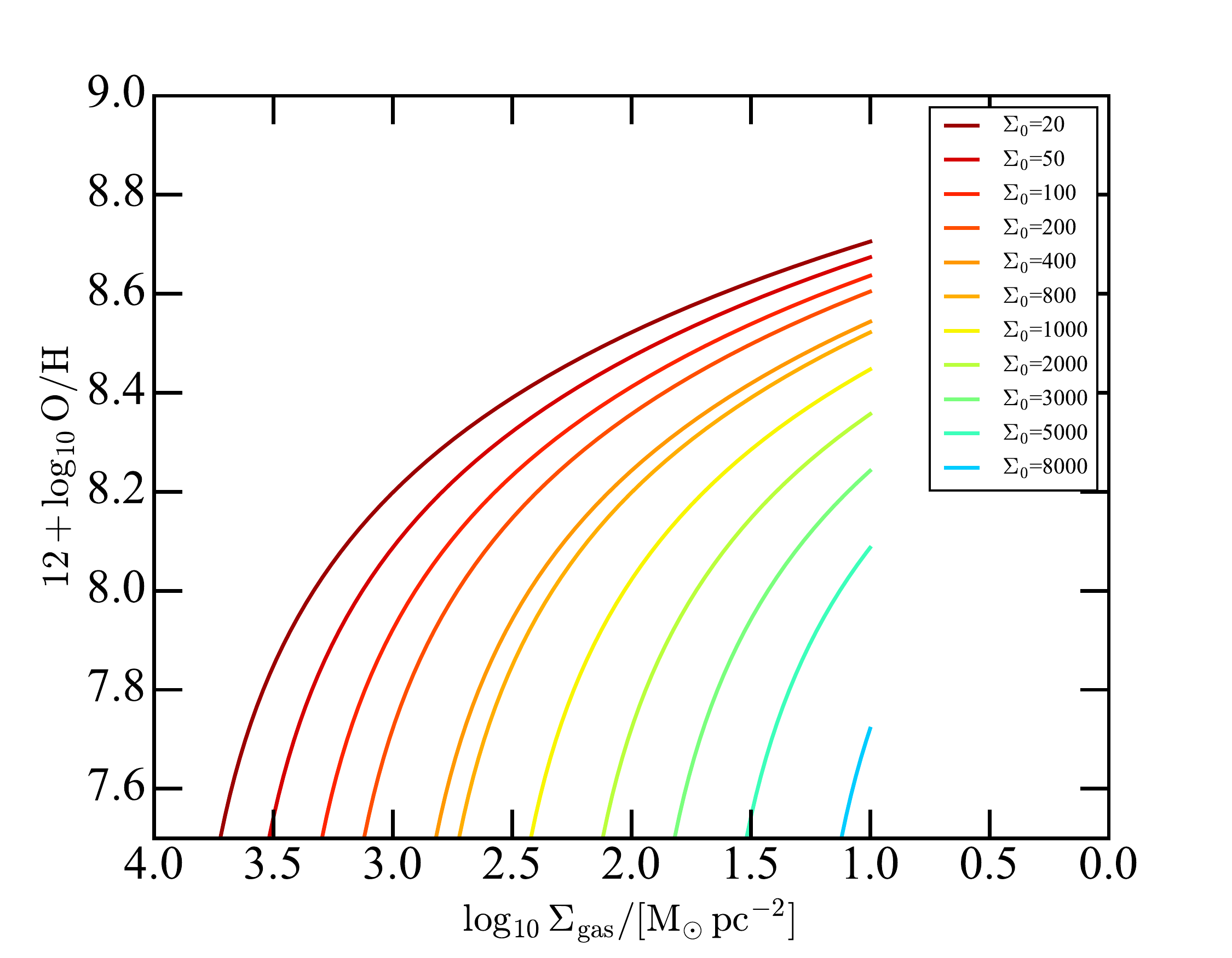}
\epsscale{1.15}
\plotone{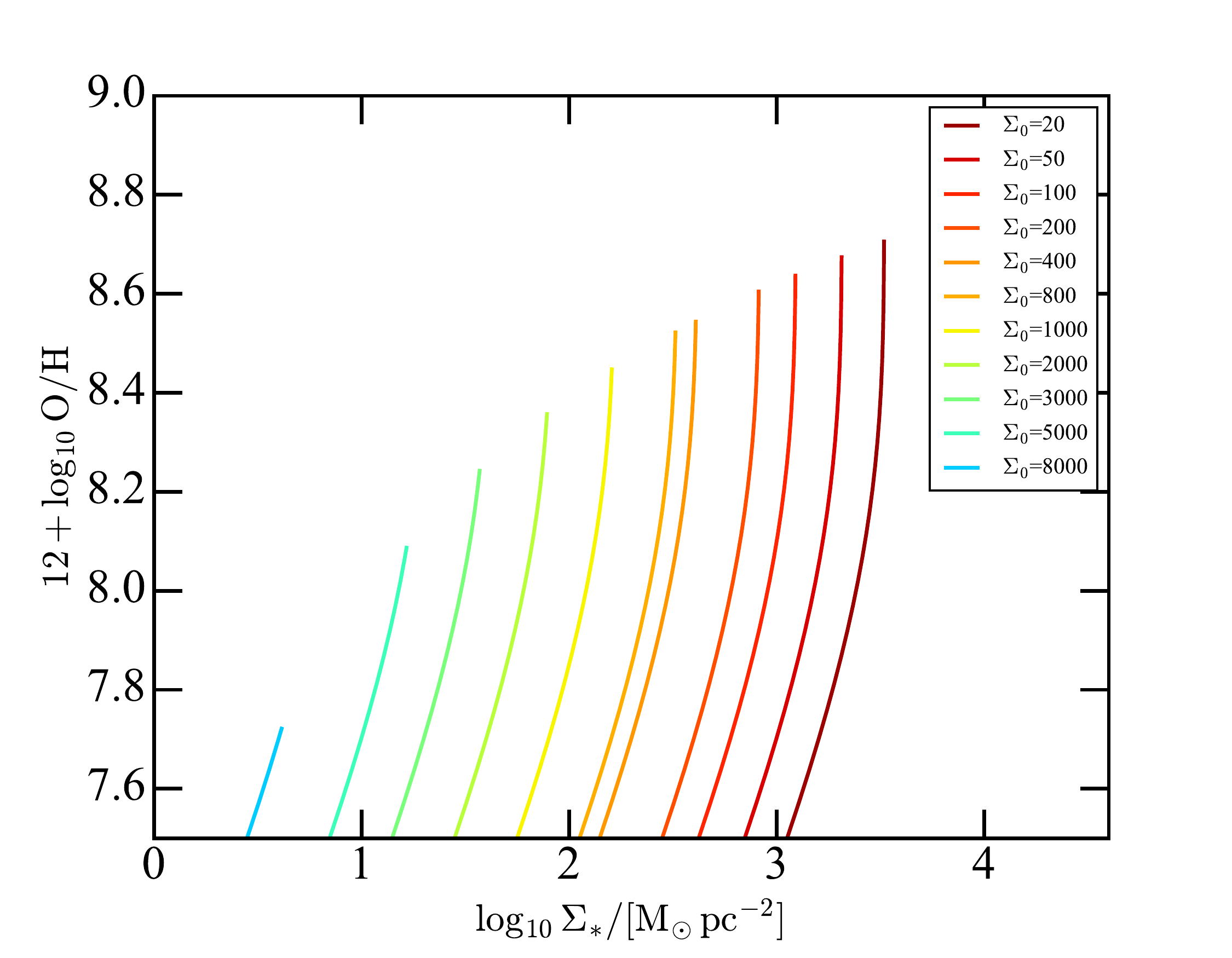}
\epsscale{1.15}
\plotone{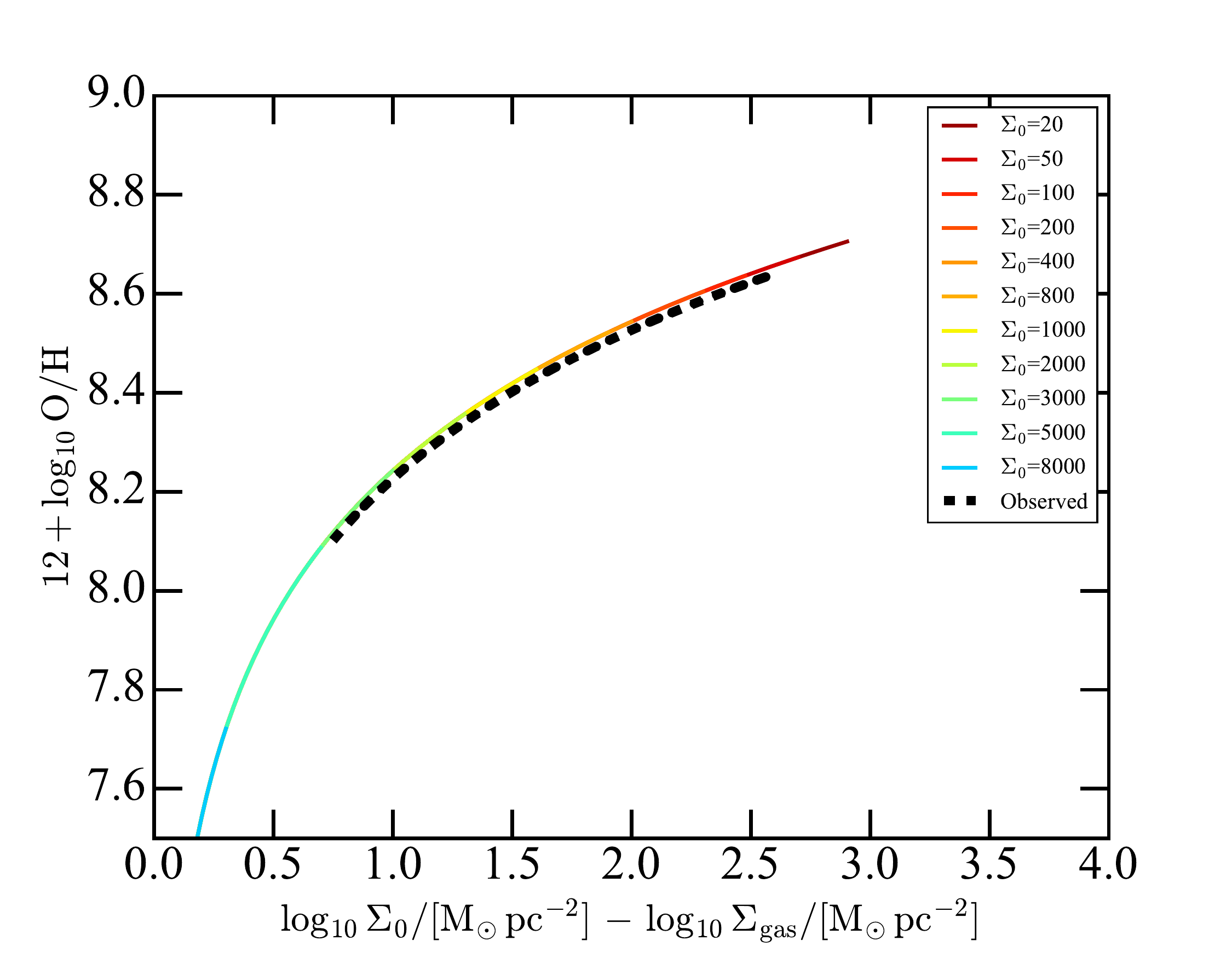}
\caption{The predicted evolutionary tracks of the local star formation history 
as a function of $\Sigma_0$ (in $\mathrm{M}_\odot\,\pc^{-2}$). 
For each track, time increases from left to right and from bottom to top, 
$\Sigma_{\rm gas}$ decreases with time, while $\Sigma_{*}$ and $Z$ increase with time.
\textit{Top}: the $\Sigma_{\rm gas}-Z$ relation. We have reversed the order of $\Sigma_{\rm gas}$ for display purposes.
\textit{Middle}: the $\Sigma_{*}-Z$ relation. 
\textit{Bottom}: the $\Sigma_{*}-\Sigma_{\rm gas}-Z$ relation, as given by Equation~\ref{eq:metalgas}.
All tracks with different $\Sigma_0$ overlap for this relation. 
The black dashed line is the same as the magenta dashed line in the right panel of Figure~\ref{fig:relation}, showing the range probed by the MaNGA survey, slightly shifted downwards for clarity.
}
\vspace{0.2cm}
\label{fig:leakymodel}
\end{figure}

The SFR surface density is related to the gas surface density through the K-S law: 
\begin{equation}
\Sigma_{\rm SFR} \equiv \frac{1}{1-\mathcal{R}}\,\frac{\ud \Sigma_{*}(t)}{\ud t} = \epsilon \Sigma_{\rm gas}^k(t) \,\mathrm{,}
\label{eq:kslaw}
\end{equation}
\noindent where $\mathcal{R}$ is the ''return fraction'', i.e.,  
the fraction of the stellar mass formed that is assumed to be instantaneously returned 
to the gas from short-lived massive stars,
and $\epsilon$ is the effective SF efficiency and $k$ is the K-S index.
Note $\epsilon$ is not unitless and its dimension depends on $k$. 
Following convention, we express $\Sigma_{*}$ and $\Sigma_{\rm gas}$ in unit of $\MSun \pc^{-2}$,
while $\Sigma_{\rm SFR}$ in unit of $\MSun \kpc^{-2}$. 
We also expect there is a threshold below which SF cannot continue, and we assume this threshold to be $10\,\MSun\pc^{-2}$ \citep[\eg][]{skillman87a, schaye04a, leroy08a}.

In global models, the outflow rate is usually assumed to be proportional to the total SFR \citep[\eg][]{springel03a, dallavecchia08a} 
and we extend this assumption to our local model.
The outflow rate is related to the SFR through
\begin{equation}
\frac{\ud \Sigma_{\rm out}(t)}{\ud t} = \eta\,\Sigma_{\rm SFR} = \frac{\eta}{1-\mathcal{R}}\,\frac{\ud \Sigma_{\rm *}(t)}{\ud t} \,\mathrm{,}
\label{eq:outflow}
\end{equation}
\noindent where $\eta$ is the mass loading factor and we assume it is constant \citep[\eg][]{springel03a, heckman15a}.

Combining the above equations gives the relation between gas consumption rate, SFR surface density, and gas surface density:
\begin{eqnarray}
\frac{\ud \Sigma_{\rm gas}(t)}{\ud t} & = & -(1+\frac{\eta}{1-\mathcal{R}}) \frac{\ud \Sigma_{\rm *}(t)}{\ud t} \\
 & = & -(1-\mathcal{R}+\eta) \epsilon \Sigma_{\rm gas}^k(t) \,\mathrm{,}
\end{eqnarray}
\noindent from which we can solve for the full star-formation history, 
including $\Sigma_{\rm gas}(t)$, $\Sigma_{*}(t)$, $\Sigma_{\rm SFR}(t)$, $\Sigma_{\rm out}(t)$, mass-weighted age of the stars, etc.
In particular, assuming $k>1$, $\Sigma_{\rm gas}(t)$ is given by
\begin{equation}
\Sigma_{\rm gas}^{1-k}(t) = \Sigma_0^{1-k} - (1-\mathcal{R}+\eta)\epsilon(1-k)(t-t_0)\,\mathrm{.}
\end{equation}

We can now derive the chemical evolution of this leaky-box model.
The metallicity ($Z\equiv\Sigma_{\rm metal}/\Sigma_{\rm gas}$) growth rate is given by
\begin{eqnarray}
\frac{\ud Z(t)}{\ud t} & = & \frac{1}{\Sigma_{\rm gas}(t)}\frac{\ud \Sigma_{\rm metal}(t)}{\ud t} - 
\frac{\Sigma_{\rm metal}(t)}{\Sigma_{\rm gas}^2(t)}\frac{\ud \Sigma_{\rm gas}(t)}{\ud t} \nonumber \\
 & = & \frac{1}{\Sigma_{\rm gas}(t)} \left( \frac{\ud \Sigma_{\rm metal}(t)}{\ud t} - Z(t)\frac{\ud \Sigma_{\rm gas}(t)}{\ud t} \right) \,\mathrm{,}
 \label{eq:metalgrowth}
\end{eqnarray}
\noindent where $\Sigma_{\rm metal}$ is the surface density of metals in the gas.
If $y$ is the total metal mass yield that a stellar population releases into the ISM normalized 
by the mass locked up in long-lived stars,
the amount of new metals that stay in the gas in the galaxy is given by the total yield minus that locked in stars 
and expelled along with outflows:
\begin{eqnarray}
\frac{\ud \Sigma_{\rm metal}(t)}{\ud t} & = & y \frac{\ud \Sigma_{*}(t)}{\ud t} - Z(t) \left(\frac{\ud \Sigma_{*}(t)}{\ud t} + \frac{\ud \Sigma_{\rm out}(t)}{\ud t}\right) \nonumber \\
 & = & \left(y-Z(t)-Z(t)\frac{\eta}{1-\mathcal{R}}\right) \frac{\ud \Sigma_*}{\ud t} \nonumber \\
 & = & \left(y-Z(t)-Z(t)\frac{\eta}{1-\mathcal{R}}\right) \times \nonumber \\
 &   & \frac{-1}{1+\eta/(1-\mathcal{R})}\frac{\ud \Sigma_{\rm gas}(t)}{\ud t}\,\mathrm{.}
\end{eqnarray}
\noindent where we have assumed the metallicity in the outflowing gas is the same as in the ISM at the time.

The metallicity growth rate is then given by
\begin{eqnarray}
\frac{\ud Z(t)}{\ud t} & = & \frac{\ud \Sigma_{\rm gas}(t)}{\Sigma_{\rm gas}(t) \ud t} \left( \frac{y}{1+\eta/(1-\mathcal{R})} \right) \,\mathrm{.}
\end{eqnarray}
Eliminating $\ud t$ gives the dependence of the metallicity on $\Sigma_0$ and $\Sigma_{\rm gas}(t)$:
\begin{eqnarray}
Z(t) - Z_0 & = & \frac{y}{1+\eta/(1-\mathcal{R})} \log \frac{\Sigma_0}{\Sigma_{\rm gas}(t)} \nonumber \\
 & = & \frac{\log(10)y}{1+\eta/(1-\mathcal{R})} \left[ \log_{10} \Sigma_0 - \log_{10} \Sigma_{\rm gas}(t) \right] \,\mathrm{.} \nonumber \\
 & & 
\label{eq:metalgas}
\end{eqnarray}
\noindent 
We have thus derived the local version of the well-known global leaky-box model of chemical evolution \citep[\eg][]{tinsley80a},
which has been used to study the global mass-metallicity relation \citep[\eg][]{zahid14a, belfiore16a}.
We assume $Z_0$ is $0.1\%$ of the solar value, though as long as it is lower than $1\%$ solar, it has no effect on any of our conclusions.

Based on the assumptions of the model (Eq.~\ref{eq:sigma0} and Eq.~\ref{eq:outflow}), we can also calculate $\Sigma_0$ as
\begin{equation}
\Sigma_0 = \Sigma_{\rm gas}(t) + \left(1+\frac{\eta}{1-\mathcal{R}}\right) \Sigma_{*}(t)\,\mathrm{,}
\end{equation}
and the metallicity can now be fully determined if we can observe $\Sigma_{*}$ and $\Sigma_{\rm gas}$ and if we know $\eta$ and $y$.
This $\Sigma_{*}-\Sigma_{\rm gas}-Z$ relation is a fundamental relation predicted by the local leaky-box model. 

Now if we assume the K-S law (Eq.~\ref{eq:kslaw}) holds and we can measure $\Sigma_{\rm SFR}$,
we can estimate the gas density $\Sigma_{\rm gas}(t)$ with 
\begin{equation}
\Sigma_{\rm gas}(t) = \left( \frac{\Sigma_{\rm SFR}(t)}{\epsilon} \right)^{1/k}\,\mathrm{.}
\end{equation}





In principle, we can constrain the parameters ($\mathcal{R}$, $y$, $\eta$, $\epsilon$, $k$) directly using the observation. 
The model, however, is non-linear and the parameters are degenerate with each other. For example, the yield $y$ and 
the loading factor $\eta$ are degenerate in the amplitude, thus a closed-box model (with $\eta=0$) with high yield 
can also fit the data well.
A robust modeling therefore requires careful treatments of the completeness (as a function of the observables).
In this first work, we choose to investigate the relation using a fiducial model with values calibrated from the literature.
In particular, we first fix the return fraction $\mathcal{R}$ to be $0.3$ for a Salpeter IMF \citep[\eg][]{tinsley80a, madau14a}.
We use $\epsilon=0.0004$ and $k=2.2$ for the K-S law in normal spiral galaxies \citep[\eg][]{misiriotis06a, bigiel08a}.
The K-S law is observed to be non-linear. For normal galaxies, the slope is $k\sim2.2$ when total gas surface density is considered, 
and is smaller ($k\sim1.2$) if only molecular gas density is included \citep[\eg][]{wong02a, boissier03a, luna06a}.  
For star-burst galaxies, the K-S law is shallower \citep[\eg][]{bigiel08a}.  
As we are interested in the total gas density for typical star-forming galaxies, we here adopt a linear K-S relation with $k=2.2$ and take the amplitude from \citet{bigiel08a}. 
For the mass loading factor $\eta$, we set it to be $1$, a choice consistent with suggestions by past studies \citep[\eg][]{martin99a, veilleux05a, schaye10a, heckman15a}.
The right panel of Figure~\ref{fig:relation} shows the observed relation with these choices. 

Fixing these three values ($\epsilon=0.0004$, $k=2.2$ and $\eta=1$), we fit the normalization for the metal yield and 
obtain $y \sim 0.003$. This yield is for oxygen ($^{16}\mathrm{O}$), 
and the total metal yield is larger by about a factor of two, $y_{\rm total} \sim 0.006$.
The values above are for a Salpeter IMF. For a Chabrier or Kroupa IMF \citep[][]{chabrier03a, kroupa01a}, 
the oxygen and total metal yield would be about $0.0045$ and $0.009$, respectively.
We plot this best-fit relation with the dashed line. 
We find it remarkable that, with these canonical values, we obtain a tight $\Sigma_{*}-\Sigma_{\rm gas}/\Sigma_{\rm SFR}-Z$ relation, 
and the fiducial model matches the observation very well.
Our best-fit metal yield is at the lower end of the theoretical estimates \citep[\eg][]{henry00a, kobayashi06a, zahid12a, vincenzo16a}.
As it is degenerate with the mass loading factor ($\eta$), if we choose a larger $\eta$, we will get a larger yield.

To take a further look at this local relation, we separate the parent spaxel samples by their galactocentric distance and the stellar mass of their host galaxy.
In Figure~\ref{fig:radialdependence}, we plot the local relation for star-forming regions outside (left) and within (right) the effective radius. 
In Figure~\ref{fig:massdependence}, we show the relation for low-mass (left) and high-mass (right) galaxies.
The dash lines in all panels are the same as in Figure~\ref{fig:relation}.
We show the best-fit local relation fits well the data of all the subsamples. 
We observe a weak dependence of the relation on the galactocentric distance and stellar mass: 
regions at larger radius and in more massive galaxies tend to be distributed above the best-fit relation with higher metallicity. 
We suspect that this weak dependence may be caused by some of the simple assumptions we made in the model: constant yield and mass loading factor, no recycled gas and metals, 
and no radial mixing. We leave detailed investigation for future work.

As similar in the global leaky-box model, given an initial gas surface density $\Sigma_0$, the leaky-box model fully describes 
the local star formation history and chemical evolution.
In Figure~\ref{fig:leakymodel}, we show for the fiducial model 
the predicted evolutionary tracks of metallicity for different $\Sigma_0$ as a function of $\Sigma_{\rm gas}$, $\Sigma_{*}$ 
and $\log_{10} \Sigma_0/\Sigma_{\rm gas}$.
Each line shows that as time increases, the metallicity and stellar surface density increase, while the gas surface density decreases.
We show that the evolution of metallicity, stellar and gas surface density, as well as their relations,
are strong functions of the initial gas surface density, while the $\Sigma_{*}-\Sigma_{\rm gas}-Z$ relation (bottom) 
does not depend on either time or $\Sigma_0$ and is a fundamental relation predicted by the local leaky-box model.

Since the local leaky-box model is fully determined by the initial gas surface density $\Sigma_0$, 
for any typical disk galaxy, if we can determine the initial surface density at the accretion time at any given radius, 
we can connect the small-scale astrophysics with the large-scale cosmological context. 
We briefly discuss how to expand the local model to a cosmological inside-out growth model in the next section. 

Some of the earlier works have presented similar ideas of localized star formation history and chemical evolution 
\citep[\eg][]{rosales12a, sanchez13a, fu13a, ho15a, carton15a, kudritzki15a}. 
In particular, \citet{ho15a} and \citet{carton15a} extended a global gas regulatory model \citep{lilly13a} by ignoring radial mass transfer,
which is also an assumption of our model, and showed that it could reproduce the radial metallicity profile for a large fraction of 
disk galaxies in their samples. They used global parameters (total stellar mass, total 
SFR) except for the metallicity in their models to reconstruct the observed density/metallicity gradient from resolved IFU 
observations. Although they did not provide a formalism for the localized star-formation history as we did,
they presented new ideas to connect the global properties of the galaxy with the local ones.
The model we suggest below outlines a way to integrate these ideas presented in their pioneering works
and our local leaky-box model to build a typical disk galaxy analytically in the cosmological context.

\section{The cosmological inside-out growth model}\label{sec:insideout}

Suppose the dark matter accretion rate of a given dark matter halo (with mass $M_{\rm DM}$) at a given time ($t$) is
\begin{equation}
\dot{M}_{\rm DM} \equiv \frac{\ud M_{\rm DM}(t)}{\ud t} = \dot{M}_{\rm DM} (M_{\rm DM}, t)\,\mathrm{,}
\end{equation}
\noindent which is a function of $M_{\rm DM}$ and $t$ and can be calibrated from simulations \citep[\eg][]{wechsler02a, correa15a}, 
the gas accretion rate (onto the galaxy) is then given by
\begin{equation}
\dot{M}_{\rm gas} (t) \equiv \frac{\ud M_{\rm gas}(t)}{\ud t} = \lambda\,f_{\rm b}\,\dot{M}_{\rm DM} (M_{\rm DM}, t)\,\mathrm{,}
\label{eq:cosmogasaccretion}
\end{equation}
\noindent where $f_{\rm b}$ is the cosmic ratio of baryon mass to dark matter and $\lambda$ is the fraction of baryons 
that fall all the way in onto the galaxy.

We assume the newly-accreted gas only stays on the outskirts and the galaxy grows from inside out. 
In this case the gas accretion rate is naturally connected to the size growth of the galaxy $\dot{R}(t)$ and the initial surface density 
at the galaxy-size radius at the accretion time $\Sigma_{0}(R)$:
\begin{eqnarray}
\dot{M}_{\rm gas} (t) &  = & n\,h(R)\,2\pi R(t)\, \frac{\ud R}{\ud t} \\
 & = & \Sigma_0(R)\,2\pi R(t)\,\dot{R}\,\mathrm{,}
\label{eq:profilesize}
\end{eqnarray}
\noindent where $n$ is the volume density when gas starts to form stars and must be closely connected to the SF density threshold for giant molecular clouds, 
$R(t)$ is the galaxy size at $t$, $h(R)$ is the initial scale height at $R$, and $\Sigma_0(R)$ is the initial total surface density at $R$.

If we can calibrate $\dot{M}_{\rm gas} (t)$ with simulations, 
we can infer the radial profile of the initial density $\Sigma_0(R)$ from the size growth of the galaxy $\dot{R}$, and vice versa.
In particular, if we know the size $R(t)$ and its growth rate $\dot{R}(t)$ of a typical disk galaxy \citep[\eg][]{vandokkum13a, vanderwel14a}, 
by applying the local leaky-box model, we can fully derive the radial profiles of $\Sigma_{\rm gas}(r,t)$, $\Sigma_{*}(r,t)$, $\Sigma_{\rm SFR} (r,t)$, 
$Z(r,t)$, and mass-weighted stellar age $\left<t_*\right> (r,t)$, where $r<R(t)$. 
IFU surveys such as CALIFA and MaNGA have started to obtain these radial profiles for a large sample of 
disk galaxies \citep[\eg][]{sanchez13a, perez13a}.
Galactic surveys, such as RAVE \citep{steinmetz06a} and APOGEE \citep{majewski16a}, have also started to provide 
chemical gradient measurements of Galactic stars \citep[\eg][]{boeche13a, hayden14a, ness16a}, 
lending support to an inside-out growth scenario for our own Milky Way.
We can also compare the relations among the above parameters and their dependence on global properties should we observe a large sample of systems, 
such as the stellar mass/SFR (in-)dependence of the $\Sigma_{*}-Z$ relation observed in our previous paper \citep[\eg][]{barrera16a} 
and the relation between global stellar mass, SFR, and central-region metallicity \citep[\eg][]{mannucci10a, laralopez10a, sanchez13a, salim14a, salim15a, bothwell16a}.
We therefore expect a full semi-analytical model can be compared with observations directly,
not only for global properties as previous-generation models, but also for local and structural properties revealed 
by IFU spectroscopic and deep high-spatial resolution imaging surveys. 
We leave the full modeling for future work.


\section{Conclusions}\label{sec:summary}

With the most recent data from the MaNGA survey, we have confirmed a tight relation between the stellar surface density, 
gas surface density, and gas-phase metallicity. 
We introduced a new local leaky-box model, in which star formation and metal production are localized 
within the same region except for the outflowing gas. 
With this model we derived closed-form solutions for the evolution of stellar surface density, gas surface density, and gas-phase metallicity, 
and showed that they follow a tight relation regardless of initial gas density and time. 
We further demonstrated that, with canonical values for the model parameters, the closed-form relation predicted by the model 
matches the observed one well. Our local leaky-box model therefore provided a natural explanation for the relationship between 
local parameters by the recent IFU observations and suggested a new look at the evolution of typical disk galaxies like our own Milky Way.
We briefly introduced how to build a cosmological semi-analytical inside-out growth model that can take into account
of the small-scale astrophysics by including the localized star formation history.

We can further refine and improve the local leaky-box model. For example, if we can observe the gas density \citep[\eg, as in the DiskMass Survey,][]{martinsson13a},
then we can investigate the local relation directly without the assumption of the Kennicutt-Schmidt law. 
The current local leaky-box model also neglects several possible effects. 
We have assumed the parameters ($\epsilon$, $k$, $\eta$, $y$) are all constant. 
In reality, the K-S index depends on $\Sigma_{\rm gas}$ \citep[\eg][]{bigiel08a}, and the mass loading factor 
must also depend on $\Sigma_{\rm SFR}$ \citep[][]{heckman15a} and also the local and/or global gravitational potential.
It is believed that radial migration of stars and gas happens on some level \citep[\eg][]{haywood08a},
though it is yet unclear how important it is in the general evolution of disk galaxies.
The expelled gas can also be recycled back to the galaxy \citep[\eg][]{oppenheimer10a, christensen16a}.
Mergers can also affect the distribution of metals \citep[\eg][]{rupke10a}.
In addition, the model we described does not address the formation and evolution of bulges and bars at the center.
It is also a statistical model and neglects structures such as spiral arms. 
We expect these open issues to be the focuses of future investigations.

On a larger scale, the outflow component can be connected to quenching due to stellar/supernova feedback.
The cosmological inside-out growth model with the localized star formation history is a natural next step of the gas regulatory model 
used for global evolution of galaxies \citep[\eg][]{bouche10a, lilly13a}. 
Instead of adding more gas to the total gas reservoir, the inside-out growth model simplifies the physical treatments 
as it adds new gas to the outskirts without interfering with the (local) reservoir on the inside. 

The MaNGA survey is continuing its operation and will provide us with six times more data by the end of the survey.
With such a large dataset, we will be able to investigate not only the local properties with IFU data themselves, 
but also the correlations between them and global properties and  large-scale structures.
Together with the rapid development of high-resolution hydrodynamical simulations and new analytical models as the one described in this paper,
we are entering a new era of galaxy formation and evolution where we can now connect directly small-scale astrophysics with the cosmological context 
in both observation and theory.

\acknowledgments

G.B.Z. acknowledges support provided by NASA through Hubble Fellowship grant \#HST-HF2-51351 awarded by the Space Telescope Science Institute, which is operated by the Association of Universities for Research in Astronomy, Inc., under contract NAS 5-26555. 
We thank an anonymous referee for many constructive comments that have helped improve this paper.

Funding for the Sloan Digital Sky Survey IV has been provided by
the Alfred P. Sloan Foundation, the U.S. Department of Energy Office of
Science, and the Participating Institutions. SDSS-IV acknowledges
support and resources from the Center for High-Performance Computing at
the University of Utah. The SDSS web site is www.sdss.org. 

SDSS-IV is managed by the Astrophysical Research Consortium for the 
Participating Institutions of the SDSS Collaboration including the 
Brazilian Participation Group, the Carnegie Institution for Science, 
Carnegie Mellon University, the Chilean Participation Group, the French Participation Group, Harvard-Smithsonian Center for Astrophysics, 
Instituto de Astrof\'isica de Canarias, The Johns Hopkins University, 
Kavli Institute for the Physics and Mathematics of the Universe (IPMU) / 
University of Tokyo, Lawrence Berkeley National Laboratory, 
Leibniz Institut f\"ur Astrophysik Potsdam (AIP),  
Max-Planck-Institut f\"ur Astronomie (MPIA Heidelberg), 
Max-Planck-Institut f\"ur Astrophysik (MPA Garching), 
Max-Planck-Institut f\"ur Extraterrestrische Physik (MPE), 
National Astronomical Observatories of China, New Mexico State University, 
New York University, University of Notre Dame, 
Observat\'ario Nacional / MCTI, The Ohio State University, 
Pennsylvania State University, Shanghai Astronomical Observatory, 
United Kingdom Participation Group,
Universidad Nacional Aut\'onoma de M\'exico, University of Arizona, 
University of Colorado Boulder, University of Oxford, University of Portsmouth, 
University of Utah, University of Virginia, University of Washington, University of Wisconsin, 
Vanderbilt University, and Yale University.

\bibliographystyle{apj}

\end{document}